\documentclass[final]{appolb}
\usepackage{epsfig}
\def\bea{\begin{eqnarray}}
\def\eea{\end{eqnarray}}
\def\beq{\begin{equation}}
\def\eeq{\end{equation}}

\begin{document}
\title{Quantum Mechanics, Random Matrices and \\BMN Gauge Theory.%
\thanks{Presented at ``Workshop on Random Geometry,'' Krakow, May 2003}%
}
\author{C.\ Kristjansen
\address{The Niels Bohr Institute \\ Blegdamsvej 17 \\ DK-2100 Copenhagen \O}}
\maketitle
\begin{abstract}
We review how the identification of gauge theory operators representing string
states in the pp-wave/BMN correspondence and their associated anomalous 
dimension reduces to the determination of the eigenvectors and the eigenvalues
of a simple quantum mechanical Hamiltonian and analyze the properties of this
Hamiltonian. Furthermore, we discuss the role
of random matrices as a tool for performing explicit evaluation of correlation
functions. 

\end{abstract}
  
\section{Introduction}
BMN gauge theory can be characterized as the theory which appears
in the lower right corner of the diagram in figure~1. 
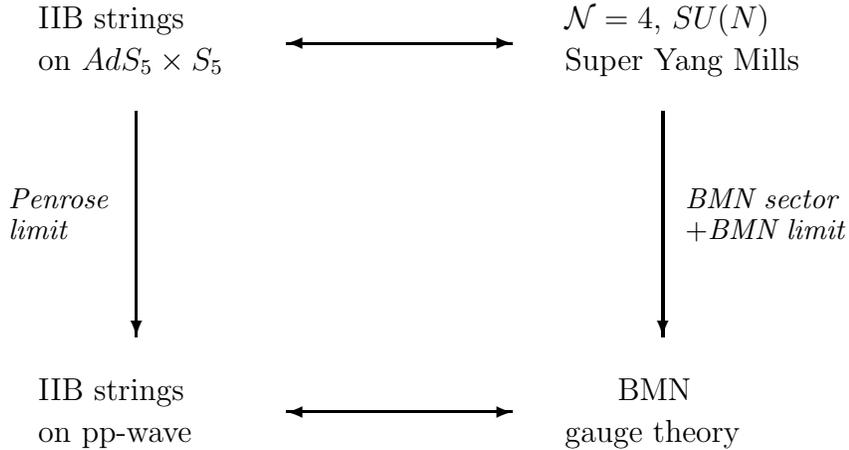
\begin{figure}
\setlength{\unitlength}{1.0cm}
\begin{picture}(12.,6.5)(0.,0.)
\thicklines
\put(1.,5.5){\large IIB strings}
\put(1.,4.95){\large on $AdS_5\times S_5$}
\put(8.,5.5){\large ${\cal N}=4$, $SU(N)$} 
\put(8.,4.95){\large Super Yang Mills}
\put(1.,.55){\large IIB strings} 
\put(1.,0){\large on pp-wave}
\put(8.7,.55){\large BMN }
\put(8.,0){\large gauge theory}
\put(4.3,.4){\vector(1,0){3}}
\put(7.3,.4){\vector(-1,0){3}}
\put(4.3,5.3){\vector(1,0){3}}
\put(7.3,5.3){\vector(-1,0){3}}
\put(2.3,4.4){\vector(0,-1){3}}
\put(0.6,3.1){\it Penrose}
\put(0.6,2.7){\it limit}
\put(9.3,4.4){\vector(0,-1){3}}
\put(9.6,3.1){\it BMN sector}
\put(9.6,2.7){+\it BMN limit}
\end{picture}
\vspace*{0.5cm}
\caption{The emergence of BMN gauge theory.}
\end{figure}
The upper line of the diagram
symbolizes the celebrated AdS/CFT conjecture which relates ${\cal N}=4$ Super
Yang Mills theory with gauge group $SU(N)$ to type IIB string theory on the
ten-dimensional geometry $AdS_5\times S_5$
~\cite{Maldacena:1997re}. Unfortunately, on this geometry
the IIB string theory has so far resisted quantization. It was understood
in~\cite{Blau:2002dy}, however,
that imposing a Penrose limit 
(left vertical arrow) on $AdS_5\times S_5$ one can
obtain a simpler ten-dimensional geometry, known as a 
pp-wave, on which
quantization of the IIB string is possible using light cone
gauge~\cite{Metsaev:2001bj,Metsaev:2002re}.
After this discovery an intriguing question was of course what
would be the corresponding procedure on the gauge theory side. The
answer was provided by Berenstein, Maldacena and Nastase who showed
that the appropriate operation consisted in considering a certain
sub-sector of the gauge theory and imposing a certain 
limit~\cite{Berenstein:2002jq} (right
vertical arrow). The theory which results from this operation is
denoted as BMN gauge theory and is via the original AdS/CFT
correspondence
conjectured to be the gauge theory dual of the IIB pp-wave string. 

As most gauge theories  ${\cal N}=4$ Super Yang Mills is 
described by two parameters the number of colours, N,
and the 't Hooft coupling constant, $\lambda=g^2_{\mbox{\tiny YM}}N$. 
The 't Hooft
coupling constant governs the field theoretical loop expansion and $N$
governs the topological expansion~\cite{'tHooft:1973jz}. As just mentioned, to
define
BMN gauge theory one has to consider a certain sub-sector of 
${\cal N}=4$ Super Yang Mills. Part of this procedure consists in 
introducing a new parameter, $J$, the $SO(2)$ R-charge. The $SO(2)$
R-charge is a charge associated with a particular $SO(2)$ sub-group
of the bosonic $SO(6)$ symmetry group of the ${\cal N}=4$ Super Yang
Mills theory. A BMN sector of ${\cal N}=4$ Super Yang
Mills can then be described as the set of operators for which the
$SO(2)$ R-charge takes some particular value, $J$, and BMN gauge theory
is obtained by, in such a sector, considering the limit 
\begin{equation}\label{BMNlimit}
J\rightarrow \infty, \hspace{0.5cm}N\rightarrow \infty,\hspace{0.7cm}
g^2_{\mbox{\tiny YM}}\mbox{ finite}, \hspace{0.5cm}
\frac{J^2}{N}\mbox{ finite}.  
\end{equation}
BMN-gauge theory is again described by two parameters
\begin{equation}\label{BMNparameter}
\lambda'=\frac{g_{\mbox{{\tiny YM}}}^2N}{J^2}, \hspace{0.7cm} \mbox{and}\hspace{0.7cm}
g_2=\frac{J^2}{N}.
\end{equation}
The parameter $\lambda'$ is an effective gauge coupling constant which governs
the loop expansion~\cite{Berenstein:2002jq} and $g_2$ is an effective genus 
counting parameter which governs the topological 
expansion~\cite{Kristjansen:2002bb,Constable:2002hw}. 

Berenstein, Maldacena and Nastase argued that in the pp-wave/BMN correspondence
string states should map to gauge invariant operators with the following 
identification of the quantum numbers
\begin{equation}
E_{l.c.}= \Delta-J.
\end{equation}
Here $E_{l.c.}$ is the light cone energy of the string state and $\Delta$ and
$J$ are respectively the conformal dimension and the $SO(2)$ R-charge of the
gauge theory operator. For a detailed discussion of the string
theory
side of the correspondence, see f.\ inst.\ the
reviews~\cite{Pankiewicz:2003pg,Plefka:2003nb}.

Operators which have definite conformal dimensions
are characterized by their two-point functions taking the form
\begin{equation}\label{confdim}
\langle {\tilde{\cal O}}_{A}(x)\bar{\tilde{\cal O}}_{B}(0)\rangle
=\frac{\delta_{AB}}{|x|^{2\Delta_A}}C_A,
\end{equation}
with $\Delta_A$ the conformal dimension and $C_A$ some normalization
constant.
In reference~\cite{Berenstein:2002jq} 
an explicit map between the states of the quantized
{\it free} IIB pp-wave string and certain operators in the dual gauge
theory was suggested. In particular, the mapping gave rise to a prediction
for the conformal dimension of these operators. 
This prediction, 
known as the BMN square root formula, re-sums the entire loop 
expansion, i.e.\
it can be expanded as an infinite power series in $\lambda'$. However, it
is limited to the case $g_2=0$, i.e.\ to the planar gauge theory. The BMN
prediction has been confirmed to all orders in 
$\lambda'$~\cite{Gross:2002su,Santambrogio:2002sb}. 
What we shall discuss
is what happens when one includes {\it non-planar} 
corrections in the gauge theory.
More precisely, we shall work at first order in $\lambda'$ and to all
orders in $g_2$.
Accordingly, we can write~eq.~(\ref{confdim}) as
\begin{equation}\label{confoneloop}
\langle {\tilde{\cal O}}_{A}(x)\bar{\tilde{\cal O}}_{B}(0)\rangle
=C_A\frac{\delta_{AB}}{|x|^{2\Delta_A^0}}
\left(1+\lambda' (\delta \Delta)_A \log |x \Lambda|^{-2}\right).
\end{equation}
Here $\Delta_A^0$ is the tree-level conformal dimension, $(\delta \Delta)_A$
is the one loop correction and $\Lambda$ is some (divergent) 
renormalization scale. One expects that the one-loop correction can be
expanded in genus as follows\footnote{In the full ${\cal N}=4$ Super
Yang Mills theory one expects an expansion similar
to eqs.~(\ref{confoneloop})
and~(\ref{deltagenus}) with $\lambda$ and $\frac{1}{N}$ 
replacing $\lambda'$ and
$g_2$. However, some one-loop anomalous dimensions have a
large-N expansion involving odd powers of $\frac{1}{N}$ and some do not
even have a well-defined double expansion in $\lambda$ and 
$\frac{1}{N}$~\cite{Beisert:2003tq}. Similar complications
have not been encountered in the BMN limit so far.}
\beq\label{deltagenus}
(\delta \Delta)_A =(\delta \Delta)_A^{(0)}+ g_2^2(\delta \Delta)_A^{(1)}+
g_2^4(\delta \Delta)_A^{(2)}+\ldots
\eeq
In our analysis we will consider only operators built from scalar fields.
The ${\cal N}=4$ Super Yang Mills theory has three complex 
scalars, $\psi$, $\phi$ and
$Z$ all in the adjoint
representation of the gauge group. By the choice of a particular
$SO(2)$ sub-group of the $SO(6)$ symmetry group of ${\cal N}=4$ Super Yang
Mills one singles out one of these three fields, say $Z$, and the $SO(2)$
R-charge is then given as the quantum number conjugate to the phase of $Z$.
For scalar operators the $SO(2)$ R-charge thus simply counts the number of
$Z$-fields. Furthermore, for such operators the tree-level conformal dimension
just equals the number of fields.
A particular set of scalar operators are the following
with tree level conformal dimension $J+2$
\begin{eqnarray}
\label{twoimpurity}
\lefteqn{\hspace{-0.5cm}\Omega_n^{J_0,J_1,\ldots,J_k}(x)= } \nonumber \\
&&\sum_{p=0}^{J_0}e^{2\pi i p n/J_0}\Tr (\phi Z^p\psi Z^{J_0-p}) \, \Tr Z^{J_1}
\ldots\Tr Z^{J_k}(x),
\hspace{0.5cm} \sum_{i=0}^k J_i= J.
\end{eqnarray}
These operators have well-defined anomalous dimensions 
in the {\it planar} BMN limit and 
correspond in that limit to direct products of 
$(k+1)$ string states, $k$ of which are vacuum
states and one is a state with two oscillators 
excited~\cite{Berenstein:2002jq}\footnote{Operators of a similar type
having well-defined planar conformal dimensions in the {\it full} gauge theory
have been constructed in reference~\cite{Beisert:2002tn}.}. 
Once one takes into
account non-planar contributions in the gauge theory, however, the operators
in eq.~(\ref{twoimpurity}) start to mix and do no longer have well-defined 
conformal dimensions. To find the gauge theory operators which correctly 
represent the string states we have to carry out a (re-)diagonalization 
process. In section~\ref{quantummechanics} we will show that
this process can be described as the process of finding the 
eigenvectors and the eigenvalues of a simple quantum mechanical Hamiltonian.
Subsequently, in section~\ref{randommatrices} we will explain how 
techniques from the field of random matrices can be used to perform 
explicit evaluation of correlation functions. The reason why random matrices
appear at all is that the propagators of the scalar fields 
in ${\cal N}=4$ Super Yang Mills
take the 
form\footnote{We use the notation and normalization of 
reference~\cite{Beisert:2002bb}.}
\bea\label{prop1}
\langle Z_{ij}(x) \bar{Z}_{kl}(0)\rangle_{free}
&=&\frac{g^2_{\mbox{\tiny YM}}}{8\pi^2 x^2}\, 
\delta_{il}\delta_{jk}\equiv\frac{g^2_{\mbox{\tiny YM}}}{8\pi^2 x^2}\,
\langle Z_{ij} \bar{Z}_{kl}\rangle, \\
\label{prop2}
\langle Z_{ij}(x) Z_{kl}(0)\rangle&=&
\langle {\bar Z}_{ij}(x) {\bar Z}_{kl}(0)\rangle=0,
\eea
and similarly for $\psi$ and $\phi$.
Here  $\langle Z_{ij} \bar{Z}_{kl}\rangle$ is easily recognized as a 
propagator
of a zero-dimensional complex matrix model
({\it cf.} section~\ref{randommatrices}). 
Thus whenever the space-time dependence of a correlator can be factored out
matrix model techniques come in handy and we sketch how such techniques make
it possible to derive exact, all genera results for certain correlation
functions~\cite{Kristjansen:2002bb,Constable:2002hw,
Okuyama:2002zn,Beisert:2002bb,Eynard:2002df}.
In the last section we briefly discuss some new insights on 
the integrability of ${\cal N}=4$
Super Yang Mills which have been obtained on the basis of the quantum
mechanical formalism~\cite{Beisert:2003tq}.

\section{The quantum mechanics \label{quantummechanics}}

Gauge theory operators which are to represent string states must be
operators with well-defined conformal dimensions. Such operators are 
characterized by being eigenstates of the dilatation operator, $\hat{D}$, 
with
eigenvalue equal to the conformal dimension. Equivalently, their
two-point functions take the canonical form in eq.~(\ref{confdim}) which at
one-loop level in BMN gauge theory looks as in eq.~(\ref{confoneloop}).
Considering a basis of operators $\{{\cal O}_{\alpha}\}$ with
identical tree-level conformal dimension, $\Delta$, a two-point function will
generically read
\beq\label{generaltwopoint}
\langle {\cal O}_{\alpha}(x)\bar{\cal O}_{\beta}(0)\rangle=
\frac{1}{|x|^{2\Delta}}\left(S_{\alpha\beta}+T_{\alpha\beta}\log|x\Lambda|^{-2}
\right),
\eeq
with $S_{\alpha\beta}$ and $T_{\alpha\beta}$ respectively a tree-level
and a one-loop mixing matrix. As pointed out in
reference~\cite{Janik:2002bd} (see also~\cite{Gross:2002mh})
it is possible to read off
the matrix elements of the dilatation operator in the basis 
$\{{\cal O}_{\alpha}\}$ from these mixing matrices.
Let us split the dilatation operator in a tree level part $\hat{D}_0$
and a one-loop correction $\delta \hat{D}$,
i.e.\
\beq
\hat{D}=\hat{D_0}+\delta \hat{D}.
\eeq
Our aim is to find linear combinations of the states $\{{\cal O}_{\alpha}\}$
which are eigenstates of the dilatation operator when the one-loop correction
is taken into account. Denoting the sought for eigenstates 
as $\tilde{{\cal O}}_A$ we have
\beq
{\cal O}_{\alpha}=V_{\alpha A} \tilde{\cal O}_A,
\label{lintrans}
\eeq
and
\beq
(\delta \hat{D})\tilde{\cal O}_A=(\delta \Delta)_A \tilde{\cal O}_A,
\eeq
as well as the (equivalent) relation~(\ref{confdim}). 
In particular in the basis
$\{{\cal O}_{\alpha}\}$ we have
\beq
(\delta \hat{D}){\cal O}_{\alpha}=
V_{\alpha A}(\delta \Delta)_AV_{A\beta}^{-1} {\cal O}_{\beta}
\equiv (\delta D)_{\alpha \beta}{\cal O}_{\beta}.
\eeq 
Inserting~eq.~(\ref{lintrans}) into 
$\langle {\cal O}_{\alpha}(x)\bar{\cal O}_{\beta}(0)\rangle$ and making use 
of eq.~(\ref{confdim}) one finds
\beq
\langle {\cal O}_{\alpha}(x)\bar{\cal O}_{\beta}(0)\rangle=
\frac{1}{|x|^{2\Delta}}
\left(V_{\alpha A}C_A V^{\dagger}_{A \beta}
+V_{\alpha A} C_A (\delta \Delta)_A V^{\dagger}_{A\beta}\log|x\Lambda|^{-2}
\right).
\eeq
Comparison to eq.~(\ref{generaltwopoint}) then gives
\beq
T_{\alpha \gamma} S^{-1}_{\gamma \beta} =(\delta
{D})_{\alpha\beta}.
\label{TSminusone}
\eeq
Let us now specialize to the following set of operators 
\begin{equation}
\label{operators}
{\cal O}_p^{J_0,J_1,\ldots,J_k}(x)=\Tr (\phi Z^p\psi Z^{J_0-p}) \, \Tr Z^{J_1}
\ldots\Tr Z^{J_k}(x), \hspace{0.7cm} \sum_{i=0}^k J_i = J.
\end{equation}
These operators all have tree level conformal dimension $J+2$. To find
the linear combinations with definite conformal dimensions at one-loop
level we must diagonalize the matrix $(\delta
{D})_{\alpha\beta}=
T_{\alpha \gamma} S^{-1}_{\gamma \beta}$. Taking into account only
planar contributions we should reproduce the BMN operators 
in~eq.~(\ref{twoimpurity}) and the associated BMN
prediction. Including non-planar corrections we go beyond BMN and
enter into the domain of {\it interacting} IIB pp-wave strings.
Due to the simple form of the scalar propagators in ${\cal N}=4$ Super
Yang Mills theory ({\it cf}.\ eqs.~(\ref{prop1}) and~(\ref{prop2}))
one can express the elements
of the tree-level mixing matrix $S_{\alpha\beta}$ as expectation
values in a zero-dimensional field theory i.e.
\beq\label{Smatrix}
S_{\alpha\beta}=\langle {\cal O}_{\alpha} \bar{\cal O}_{\beta}\rangle,
\eeq 
where ${\cal O}_{\alpha}$ and $\bar{\cal O}_{\beta}$ now consist of 
space-time
independent fields and contractions are carried out using the Feynman
rules
\beq\label{contract}
\langle Z_{ij} \bar{Z}_{kl}\rangle=\delta_{il}\delta_{jk},
\hspace{0.7cm} \langle Z Z\rangle=\langle \bar{Z}\bar{Z}\rangle=0,
\eeq
and similarly for $\psi$ and $\phi$. The relation~(\ref{Smatrix}) defines
an inner product on the space of states that we shall denote as the gauge
theory inner product.
Moreover, as shown in reference~\cite{Beisert:2002bb,Beisert:2002ff}, 
the matrix
elements $T_{\alpha\beta}$ for operators involving only scalar fields
can be expressed in an analogous manner by means of an effective
Hamiltonian, $\hat{H}$. More precisely, one has
\beq
\label{Tmatrix}
T_{\alpha\beta}=\langle{\cal O}_{\alpha}\, \hat{H}\, 
\bar{\cal O}_{\beta}\rangle,
\eeq
where the notation and the contraction rules are as above and where
$\hat{H}$ encodes the combinatorial structure of the ${\cal N}=4$ Super Yang
Mills interaction
\begin{equation}\label{H}
\hat{H}=-
\frac{g_{\mbox{\tiny YM}}^2}{8 \pi^2}
\,\,:\left(\Tr [\bar{Z},\bar{\phi}][Z,\phi]
+\Tr [\bar{Z},\bar{\psi}][Z,\psi]+\Tr [\bar{\phi},\bar{\psi}][\phi,\psi]
\right):.
\end{equation}
Here the normal ordering means that contractions between two fields of
$\hat{H}$ are forbidden. Notice that the operator $\hat{H}$ is Hermitian
with respect to the gauge theory inner product.  
Unlike believed until recently,
determining the matrix elements  $(\delta
{D})_{\alpha\beta}=
T_{\alpha \gamma} S^{-1}_{\gamma \beta}$ does not require determining
neither $S$ nor $T$ as we shall now explain. In
evaluating~(\ref{Tmatrix}) one can perform the contractions in any
convenient order. In particular, one may start by contracting $\hat{H}$ with
${\cal O}_{\alpha}$. In doing so for an operator of the type~(\ref{operators}) 
one observes that the contraction produces a linear combination of
operators of the same type, i.e.\
\begin{equation}
\hat{H}{\cal O}_\alpha=H_{\alpha\gamma}\, {\cal O}_\gamma.
\label{expansion}
\end{equation}
Thus, we have
\begin{equation}\label{Hcontract}
T_{\alpha\beta}=\langle(\hat{H}{\cal O}_{\alpha})\, 
\bar{\cal O}_{\beta}\rangle 
=H_{\alpha\gamma} \langle{\cal O}_{\gamma}\, \bar{\cal O}_{\beta}\rangle 
=H_{\alpha\gamma} S_{\gamma\beta},
\end{equation}
or comparing to~(\ref{TSminusone})
\beq
H_{\alpha\beta}=T_{\alpha\gamma}S^{-1}_{\gamma\beta}
=(\delta D)_{\alpha\beta}.
\eeq
Hence, to determine the matrix elements of $\delta \hat{D}$ we only need to
determine the expansion coefficients $H_{\alpha\gamma}$ in 
eq.~(\ref{expansion}). The matrix  
$H_{\alpha\gamma}$ is obviously not Hermitian. 
However, it is related to its Hermitian
conjugate by a similarity transformation, i.e.\
\beq
H^{\dagger}_{\alpha\beta}\equiv H^{*}_{\beta\alpha}
=S^{-1}_{\alpha\gamma}H_{\gamma\delta}S_{\delta\beta},
\eeq
which can be seen by first contracting $\hat{H}$ with $\bar{\cal O}_{\beta}$
in eq.~(\ref{Hcontract}).
This in particular implies that $\delta\hat{D}$ has real eigenvalues as we
expect. Furthermore, the eigenvectors of $H_{\alpha\beta}$
corresponding to different eigenvalues are automatically orthogonal
with respect to the gauge theory inner product.

Applying the operator $\hat{H}$ in equation~(\ref{H}) to a state of the
type~(\ref{operators}) using the contraction rules~(\ref{contract}) one
finds that $\hat{H}$ can conveniently be split into three parts, a
trace-conserving one $\hat{H}_0$, a trace-increasing one $\hat{H}_+$
and a trace-decreasing one $\hat{H}_-$, i.e.
\beq
\hat{H}=-\frac{g^2_{\mbox{\tiny YM}}N}{4 \pi^2}
\left(\hat{H}_0+\frac{1}{N}\hat{H}_++\frac{1}{N}\hat{H}_-\right),
\eeq 
where
\beq\label{H0}
\hat{H}_0 {\cal O}_p^{J_0,J_1,\ldots,J_k}
={\cal O}_{p+1}^{J_0,J_1,\ldots,J_k}
-2{\cal O}_{p}^{J_0,J_1,\ldots,J_k}+{\cal O}_{p-1}^{J_0,J_1,\ldots,J_k},
\eeq
takes the form of a discrete second derivative and where the expressions for 
$\hat{H}_+{\cal O}$ and $\hat{H}_-{\cal O}$ can be found in
reference~\cite{Beisert:2002ff}\footnote{The trace-conserving part 
$\hat{H}_0$,when acting on a general single trace operator of scalar fields
has been identified as a Hamiltonian of an integrable $SO(6)$ spin 
chain~\cite{Minahan:2002ve}.}.
Starting from the discrete Hamiltonian we can derive a continuum,
i.e.\ a BMN version. To do so we must consider $J\rightarrow \infty$,
$N\rightarrow \infty$ while keeping fixed
$\lambda'=\frac{g^2_{\mbox{\tiny YM}}N}{J^2}$ and $g_2=\frac{J^2}{N}$
({\it cf.} eqs.~(\ref{BMNlimit}) and~(\ref{BMNparameter})). 
Preparing for this we introduce
the quantities 
\beq
r_i=\frac{J_i}{J}, \hspace{0.5cm} i\in \{0,1,\ldots,k\}, \hspace{0.5cm}
\sum_{i=0}^k r_i=1,\hspace{0.5cm} x=\frac{p}{J_0}\leq r_0.
\eeq
Then we imagine that all the $J_i$ become very large so that we can
view the $r_i$ and $x$ as continuous variables. Accordingly, we
replace our discrete set of operators in eq.~(\ref{operators}) with a
continuum set of states, i.e.
\beq
{\cal O}_p^{J_0,J_1,\ldots,J_k}\longrightarrow |x;r_1,\ldots,r_k\rangle,
\eeq
where the order of the 
$r$-quantum numbers is unimportant. Finally,
imposing the BMN limit in the explicit expression for $\hat{H} {\cal O}$
one finds
\beq
\hat{H}\longrightarrow \frac{\lambda'}{4 \pi^2} h,
\hspace{0.7cm} h=h_0+g_2(h_++h_-), \label{h}
\eeq
where obviously the discrete second derivative in~(\ref{H0}) turns
into a continuous one, i.e.
\beq
h_0=-\partial_x^2,
\eeq
and where $h_+$ and $h_-$ become more involved integro-differential
operators ({\it cf}.\ reference~\cite{Beisert:2002ff}). It is important to
notice that the expression~(\ref{h}) is exact at one-loop order. In
the planar limit ($g_2=0$) we have $h=h_0$. All information about 
higher genera contributions is encoded in a single trace-splitting
term $g_2h_+$ and a single trace-joining term $g_2h_-$. In particular,
there are no
terms of higher order in $g_2$.
The continuum
Hamiltonian not allowing for an exact diagonalization it becomes
natural to split $h$ into a free part $h_0$ and a perturbation 
with $g_2$ playing the role of a perturbation
parameter 
\beq
h=h_0+g_2 V,\hspace{0.7cm} V=h_++h_-.
\eeq
The free Hamiltonian $h_0$ is easily diagonalized. Due to
the cyclicity of the trace the variable $x$ is effectively periodic and
the eigenstates of $h_0$ are simply the Fourier modes 
\beq
\label{fourier}
|n;r_1, \ldots, r_k \rangle = \frac{1}{\sqrt{r_0}}\,
\int_0^{r_0} dx\, e^{2\pi i n x/r_0}\, 
|x;r_1, \ldots, r_k \rangle,\hspace{0.7cm} n\in {\cal Z}.
\eeq
The corresponding eigenvalues read
\beq
E_{|n;r_1,\ldots,r_k\rangle}^{(0)}=4 \pi^2 \frac{n^2}{r_0^2}.
\eeq
We notice that in the planar limit we, as promised, 
reproduce the BMN operators in eq.~(\ref{twoimpurity})
up to normalization. In the basis given by~(\ref{fourier}) the perturbation
acts as follows~\cite{Beisert:2002ff}
\begin{eqnarray}
\label{h-on-eigen}
\lefteqn{
h_+\,  |n;r_1, \ldots, r_k \rangle =\nonumber }\\  
&&
\int_0^{r_0} dr_{k+1}\, \sum_{m=-\infty}^{\infty}
\,\,\frac{4 m \sin^2\left(\pi n\frac{r_{k+1}}{r_0}\right)}
{\sqrt{r_0}\sqrt{r_0-r_{k+1}}\left(m-n \frac{r_0-r_{k+1}}{r_0}\right)}
~|m;r_1, \ldots, r_{k+1} \rangle ,
\nonumber \\
\lefteqn{ h_-\,  |n;r_1, \ldots, r_k \rangle = \nonumber}\\ 
&&
\sum_{i=1}^k\; \sum_{m=-\infty}^{\infty}
\,\,\frac{4\, r_i\, m \sin^2\left(\pi m\frac{r_i}{r_0+r_i}\right)}
{\sqrt{r_0}\sqrt{r_0+r_i}\left(m-n \frac{r_0+r_i}{r_0}\right)}
~|m;r_1, \ldots, 
\makebox[0pt]{\,\,$\times$}r_{i},\ldots,r_k \rangle, 
\end{eqnarray} 
where the quantity $r_0$ refers to the state on the left hand side
of the equations.
We can now proceed by quantum mechanical perturbation theory to evaluate
order by order in $g_2$ the non-planar
corrections to our eigenstates and eigenvalues
(i.e.\ to the gauge theory operators dual to string states and their 
associated conformal dimensions). For that purpose it is convenient to
to introduce an inner product on the space of states given
in~(\ref{fourier})
\begin{equation}
\langle n;s_1,\ldots s_l|m;r_1,\ldots,r_k\rangle 
=\delta_{kl}\,\delta_{mn}
\sum_{\pi\in S_k}
\prod_{i=1}^k\, \delta(s_i-r_{\pi(i)}),
\label{newinnerp}
\end{equation}
where the sum runs over permutations of $k$ elements.
This inner product is only a computational tool which makes it
possible to represent calculations in the usual language of quantum
mechanics. 

As is well-known, the first order energy shift in quantum
mechanical perturbation theory is given by the diagonal elements of
the perturbation --- provided there are no degeneracies or matrix elements
between degenerate states vanish. Our
perturbation is entirely off-diagonal but we have huge degeneracies.
For simplicity, let us consider the case of a single trace state
$|n\rangle$. Such a state is degenerate with a multi-trace
state $|m;r_1,\ldots,r_k\rangle$ if 
$n=\pm \frac{m}{1-(r_1+\ldots+r_k)}$~\cite{Constable:2002vq}. 
The
perturbation $(h_++h_-)$ 
can at worst have non-vanishing matrix elements between
states for which the number of traces differs by one. However, from
the explicit form of the matrix elements in eq.~(\ref{h-on-eigen}) we
see that such matrix elements vanish for  degenerate
states.\footnote{Notice that this statement is only true in the BMN
limit and not in the full ${\cal N}=4$ Super Yang Mills theory ({\it cf.}\
reference~\cite{Beisert:2003tq}).} Thus, we {\it can} actually use the
formulas from first order 
non-degenerate perturbation theory and conclude that
there is no energy shift for the state $|n\rangle$ at this order but
that the state itself gets corrected through mixing with double trace
states (that are not degenerate with $|n\rangle$). Since the
degeneracies
are not lifted at leading order in perturbation theory we also have to 
worry about these at next to leading order. 
Defining
\beq
{\cal P}_{|\alpha\rangle}=\frac{1-|\alpha \rangle \langle \alpha|}
{E_{|\alpha\rangle}^{(0)}-h_0},
\eeq
the familiar formulas for
the second order correction to energies and states in {\it
non-degenerate} perturbation theory read
\beq\label{energy2}
E_{|\alpha\rangle}^{(2)}=
\langle \alpha|V\,{\cal P}_{|\alpha\rangle}
V|\alpha\rangle
=
\sum_{\beta \neq \alpha}
\frac{|\langle \alpha|V|\beta\rangle|^2}
{E^{(0)}_{|\alpha\rangle}-E^{(0)}_{|\beta\rangle}}, \
\eeq
and 
\beq
\label{state2}
|\alpha\rangle^{(2)}=
{\cal P}_{|\alpha\rangle} V {\cal P}_{|\alpha\rangle} V |\alpha\rangle=
\sum_{\beta \neq \alpha}|\beta\rangle \frac{\langle \beta|V 
{\cal P}_{|\alpha\rangle} V 
 |\alpha\rangle}{E^{(0)}_{|\alpha\rangle}-E^{(0)}_{|\beta\rangle}},
\eeq
where the last expression holds for an off-diagonal perturbation.
Naively
applying the formula~(\ref{energy2}) to our 
state $|n\rangle$ the only
non-vanishing contributions come from intermediate
double trace states of the type
$|m;r\rangle$ which are not degenerate with $|n\rangle$,
the sum is finite and gives the value presented in 
references~\cite{Beisert:2002bb,Constable:2002vq}. However,
applying the formula~(\ref{state2}) to our state
$|n\rangle$ we encounter a divergence if the matrix element
$\langle m;r_1,r_2|V{\cal P}_{|n\rangle}V|n\rangle$ is 
non-vanishing for $n=\pm\frac{m}{1-r_1-r_2}$ or if 
$\langle -n|V{\cal P}_{|n\rangle}V|n\rangle \neq 0$.
The  possibility
of such a divergence was first discussed in
reference~\cite{Constable:2002vq}. In the present formalism it is
simple to evaluate the problematic matrix elements and what one finds
is that the latter vanishes whereas the former
is {\it non-vanishing} for $n=+\frac{m}{1-r_1-r_2}$~\cite{Beisert:2002ff}. 
Thus
non-degenerate perturbation theory can in general not be applied to
the state $|n\rangle$. The only case for which it remains valid
is the case $n=1$ where degeneracy with multi-trace states is 
excluded~\cite{Constable:2002vq}. 
To correctly find the
second order energy shift to the state $|n\rangle$ we have to 
diagonalize the operator
\beq
\hat{M}=V {\cal P}_{|n\rangle} V,
\eeq
in the space of states degenerate with $|n\rangle$, see f.\ 
inst.~\cite{Messiah}. 
There will be
non-vanishing matrix elements between $(2k+1)$ and $(2k+3)$ states 
for all $k$ as 
well as non-vanishing matrix elements connecting $k$-trace  states with
$k$-trace states. So far it has not been possible to carry out the 
required diagonalization. 
The breakdown of non-degenerate perturbation theory was
interpreted in~\cite{Freedman:2003bh} on the string theory side
as an instability 
causing a single string state to decay into degenerate triple string
states. 

\section{Random matrices \label{randommatrices}}
The role matrix models play in with BMN gauge theory is similar
to the role they play in the study of 2D quantum gravity, see f.\ inst.\
~\cite{Ambjorn}, (and in the Dijkgraaf-Vafa approach to  
supersymmetric
gauge theories~\cite{Dijkgraaf:2002xd}). They constitute a
convenient tool for handling
the combinatorics of Feynman diagrams with trivial space-time dependence. 

Consider a two-point function of operators built from scalar fields. Due to
the simple form of the propagators, {\it cf.} eqs.~(\ref{prop1})
and~(\ref{prop2}), (or due to
conformal invariance), at tree level one can immediately factor out the 
space-time dependence and one is left with a correlation function in a 
zero-dimensional field theory. The same is true in the case of three-point
functions.  Furthermore, as a consequence of supersymmetry, some operators
have two- and three-point functions which are protected, i.e.\
which do not get 
any quantum corrections. Examples of such operators are BMN operators with
zero or one impurity, i.e.
\bea
\Omega^{J_1,\ldots,J_k}(x)&=&\Tr Z^{J_1}\ldots \Tr Z^{J_k} (x),
\hspace{0.7cm}\sum_{i=1}^k J_i=J, \\
\Omega_0^{J_0,J_1,\ldots,J_k}(x)&=&\Tr \phi Z^{J_0}
\Tr Z^{J_1}\ldots \Tr Z^{J_k} (x),
\hspace{0.7cm}\sum_{i=0}^k J_i=J,
\eea
or two-impurity operators with mode-number $n=0$, {\it cf.}
eq.~(\ref{twoimpurity}). For these operators, which correspond to
supergravity states on the string theory side, tree-level two- and
three-point functions are thus exact and can conveniently be obtained
using techniques from the field of random matrices. As an example,
let us consider the following two-point function
\beq\label{Om1}
\langle \Omega^J(x) \bar{\Omega}^{J_1,\ldots,J_k}(0)
\rangle =
\left(\frac{g^2_{\mbox{{\tiny YM}}}}{8 \pi^2 x^2}\right)^J
\langle \Tr Z^J \prod_{i=1}^k \Tr \bar{Z}^{J_i}\rangle.
\eeq
Here we have factored out the trivial space-time dependence and the
remaining expectation value is to be evaluated using the contraction 
rules~(\ref{contract}). As we are instructed to take $J\rightarrow
\infty$ in the BMN limit the combinatorics of these contractions
would be very involved were it not for the existence of
matrix model techniques. We can represent the expectation value above 
as the following matrix integral
\beq
\langle
\Tr Z^J \prod_{i=1}^k \Tr \bar{Z}^{J_i}
\rangle=
\int d Z d\bar{Z} \exp\left(-\Tr\bar{Z} Z\right) \Tr Z^J \prod_{i=1}^k
\Tr \bar{Z}^{J_i},
\label{matrixexp}
\eeq
with  measure 
\beq
dZ d\bar{Z}=\prod_{i,j=1}^{N} \frac{d \mbox{Re} Z_{ij}\, d \mbox{Im}
 Z_{ij}}{\pi},
\eeq
as the Gaussian term produces exactly the contraction rule in 
eq.~(\ref{contract}).
A matrix integral
like the one in eq.~(\ref{matrixexp}) can be evaluated using an old
method due to Ginibre~\cite{Ginibre}, see
also~\cite{Mehta}. Diagonalizing $Z$ by a similarity transformation
i.e.\ writing
\beq\label{similarity}
Z= X D X^{-1},
\eeq
where $X$ as well as $D$ are complex matrices and $D$ is diagonal it
becomes possible
to integrate out the non-diagonal degrees of freedom~\cite{Ginibre}. 
Doing so
leaves one with an integral over only diagonal degrees of freedom which
can be evaluated {\it
exactly}~\cite{Beisert:2002bb,Okuyama:2002zn}. The result reads
\bea
\lefteqn{\hspace{-0.7cm}\langle \Tr Z^{J} \prod_{i=1}^k 
\Tr \bar{Z}^{J_i}\rangle }
\label{exact} \\
&=&\frac{1}{J+1}
\left\{
\frac{\Gamma(N+J+1)}{\Gamma(N)}-
\sum_{i=1}^k \frac{\Gamma(N+J-J_i+1)}{\Gamma(N-J_i)}
\right.
\nonumber  \\
&&
+\sum_{1\leq i_1<i_2\leq k} \frac{\Gamma(N+J-J_{i_1}-J_{i_2}+1)}
{\Gamma(N-J_{i_1}-J_{i_2})}-\cdots
\left.
+(-)^k\frac{\Gamma(N+1)}{\Gamma(N-J)}\right\}. \nonumber
\eea
We stress that since the correlation function~(\ref{Om1}) is known to
be protected we have hereby determined its value to all orders in the 
loop expansion and to all genera. A similar statement of course holds
in the BMN limit where we get
\beq
\frac{1}{J\, N^J} \langle \Tr Z^{J} \prod_{i=1}^k \Tr \bar{Z}^{J_i}\rangle
\longrightarrow  g_2^{k-1}\,
\prod_{i=1}^k \frac{\sinh\left(\frac{g_2\, r_i }{2}\right)}
{\frac{g_2}{2}}.
\eeq
The method of Ginibre allows one to evaluate (in principle) any 
 expectation value involving a product composed of factors of the
type $\Tr Z^{J_i}$ and $\Tr \bar{Z}^{K_i}$. This type of expectation
value is also accessible by
character expansion~\cite{Kostov:1996bs}. 
Using either approach one can obtain exact 
expressions for a large number of protected two- and three-point
functions, f.\ inst.\
\beq\label{Omzero}
\langle \Omega_0^J(x) \bar{\Omega}_0^{J_0,J_1,\ldots,J_k}(0)\rangle
=\left (\frac{g^2_{\mbox{\tiny YM}}}{8 \pi^2 x^2}\right) \frac{1}{J+1}
\langle \Omega^{J+1}(x) \bar{\Omega}^{J_0+1,J_1,\ldots,J_k}(0)\rangle,
\eeq
where to arrive at~(\ref{Omzero}) we have contracted by hand the two
impurities. 

In general correlation functions of scalar BMN operators are not protected.
Of course, at {\it tree level} any two- or three-point function of
such operators can be reduced to an expectation value in a
zero-dimensional
gaussian one-matrix model, the strategy being the same as above:
One first factors out the trivial space-time dependence and next 
contracts by hand the finite number of impurities. 
However, not all the resulting matrix model expectation
values can be evaluated by the method of Ginibre or by character
expansion. As an example, let us consider a tree-level
two-point function of
operators of the type appearing in~eq.~(\ref{twoimpurity})
\begin{eqnarray}\label{Omnm}
\lefteqn{\hspace{-0.5cm}\langle \Omega_n^J(x) \bar{\Omega}_m^{J}(0)\rangle
=}\nonumber \\
&&
\left (\frac{g^2_{\mbox{\tiny YM}}}{8 \pi^2 x^2}\right)^{J+2}
\sum_{p,q=0}^J e^{2\pi i(n p-mq)/J}
\langle \Tr(Z^p \bar{Z}^q) \Tr(Z^{J-p}\bar{Z}^{J-q})\rangle.
\end{eqnarray}
Clearly the product of traces in eq.~(\ref{Omnm}) is not diagonalized by
the similarity transformation~(\ref{similarity})\footnote{There exists
another possibility for diagonalizing a complex matrix, namely writing
$Z=V {\cal D} W^{\dagger}$  with $V$ and $W$
unitary and ${\cal D}$ diagonal. 
Exploiting this on can obtain expectation values of products
of traces of the form $\Tr (\bar{Z}Z)^k$ and that even for a general
$U(N)\times U(N)$ invariant potential~\cite{Morris:1990cq,Ambjorn:1992xu}, 
but also not this
method applies to a correlation function like the one in
eq.~(\ref{Omnm}).}. However, there exists a strategy by means of which
one can evaluate any expectation value of traces of words of $Z$'s and
$\bar{Z}$'s order by order in the genus expansion, namely a strategy
based on loop equations~\cite{Eynard:2002df}. Loop equations express
the invariance of the matrix model partition function under certain
analytical re-definitions of the integration variables. These equations
are most conveniently expressed in terms of generating functions. For the
correlator in eq.~(\ref{Omnm}) the relevant generating function is
\begin{equation}
W(x_1,y_1;x_2,y_2)=\left<
\Tr\left( \frac{1}{x_1-Z}\frac{1}{\bar{y}_1-\bar{Z}}\right)
\Tr\left( \frac{1}{x_2-Z}\frac{1}{\bar{y}_2-\bar{Z}}\right)\right>.
\end{equation}
This function fulfills
\begin{eqnarray}\label{mixmatrix}
\lefteqn{\
W(X e^{\frac{-i\pi n}{J}},X e^{\frac{-i\pi m}{J}};
Xe^{\frac{i\pi n}{J}},
Xe^{\frac{i\pi m}{J}}) = }\\
&& e^{i \pi (m-n)}\,
\sum_{J=0}^\infty (X\bar{X})^{-J-2} \sum_{p,q=0}^J 
\left< \Tr (Z^{J-p} \bar{Z}^{J-q}) \Tr (Z^p \bar{Z}^q) \right>
e^{2 i\pi(n p -m q)/J}, \nonumber
\end{eqnarray}
and allows us to easily recover the sum in~eq.~(\ref{Omnm}) by a contour
integration (over $X\bar{X}$). The result of the contour integration 
depends on the analyticity structure of $W(x_1,y_1;x_2,y_2)$. It can
be shown that $W(x_1,y_1;x_2,y_2)$ only has singularities in the form
of poles~\cite{Eynard:2002df}. For the choice of arguments of $W$ in
eq.~(\ref{mixmatrix}) the position and the order of the poles, not
surprisingly, depend on $n$ and $m$. Taking this into account one finds
in the BMN limit at tree level and to genus 
one~\cite{Kristjansen:2002bb,Constable:2002hw,Eynard:2002df}
$$
\langle \Omega_n^J(x) \bar{\Omega}_m^J(0) \rangle
\longrightarrow \left( \frac{g^2_{\mbox {\tiny YM}}}{8 \pi^2
x^2}\right)^{J+2}
S_{nm},\hspace{0.7cm} S_{nm}=
\delta_{nm}+g_2^2 M_{nm}+{\cal O}(g_2^4), \nonumber
$$
where $S_{nm}$ was defined (more generally) in
eq.~(\ref{generaltwopoint})
and where
$$
M_{n m}=M_{m n}=
\cases{\frac{1}{24} & $n=m=0$\cr
0& $n\neq0,m= 0$ \cr
\frac{1}{60}
-\frac{1}{24\pi^2n^2}
+\frac{7}{16\pi^4n^4}& $n=m$ and $n\neq0$
\cr
\frac{1}{48\pi^2n^2}+\frac{35}{128\pi^4n^4}& $n=-m$ and $n\neq0$ \cr
\frac{2\pi^2(n- m)^2-3}{24( n-m)^4\pi^4} +
\frac{2\,  n^2-3\, n\, m+2\,  m^2}
{8\, n^2\,  m^2\,( n- m)^2\,\pi^4}
& $|n|\neq |m|$ and 
$n\neq0\neq m$ 
\cr}\nonumber
$$
The genus two result can be found in
references~\cite{Constable:2002hw,Eynard:2002df}.
One can of course also, by means of the effective vertex $H$ express
the one-loop correction to non-protected two-point functions, i.e.\
the quantity $T_{\alpha\beta}$, as a matrix model expectation value.
As stressed earlier, when aiming at determining conformal dimensions
one has no need of knowing neither $S_{\alpha\beta}$ nor
$T_{\alpha\beta}$. The quantity $S_{\alpha\beta}$ nevertheless
has an interesting
interpretation on the string theory side as it is conjectured to
provide the transition between gauge theory operators and string
states encompassing the effects of string 
interactions~\cite{Vaman:2002ka,Pearson:2002zs,Spradlin:2003bw}.
\section{Conclusion}

Matrix model techniques played an important role in the early investigations
of BMN gauge theory leading to the discovery of the genus counting parameter
$g_2$ and allowing, via the use of effective vertices, for the first
calculations of higher genus corrections to the one-loop anomalous
dimension of BMN
operators~\cite{Kristjansen:2002bb,Constable:2002hw,Beisert:2002bb,Constable:2002vq}.
Later, it was understood that by focusing on the dilatation operator
of the ${\cal N}=4$ Super Yang Mills theory these calculations could
be considerably simplified, being equivalent to the diagonalization of a
simple quantum mechanical Hamiltonian~\cite{Beisert:2002ff}. The
quantum mechanical picture applies also
to the full ${\cal N}=4$ Super Yang Mills theory 
and this has revealed a very promising and yet to be explored
underlying
integrability structure~\cite{Beisert:2003tq}. Integrability at the
planar
one-loop level was established in~\cite{Minahan:2002ve} for scalar
operators and was recently generalized to all operators
in~\cite{Beisert:2003jj,Beisert:2003yb}. The study
of~\cite{Beisert:2003tq} provided evidence for two-loop integrability
and
lead to the conjecture that integrability would hold at all
loop orders. 

\vspace{6mm}
\noindent
{\bf Acknowledgments} 

\noindent
I thank my collaborators Niklas Beisert, Bertrand Eynard, Jan Plefka,
Gordon Semenoff and Matthias Staudacher for many inspiring and
enjoyable discussions and other events.

\end{document}